\documentclass[aip,floatfix,reprint,groupedaddress]{revtex4-1}

\usepackage{graphicx} 
\usepackage{dcolumn}
\usepackage{amsfonts}
\usepackage{amssymb}
\usepackage{amsmath}
\usepackage{bm}
\usepackage{epstopdf}
\newcommand{\be}{\begin{equation}}
\newcommand{\ee}{\end{equation}}
\newcommand{\bea}{\begin{eqnarray}}
\newcommand{\eea}{\end{eqnarray}}

\begin{document} 

\title{Trapping ultracold gases near cryogenic materials with rapid reconfigurability}

\author{Matthew A. Naides}
\author{Richard W. Turner}
\author{Ruby A. Lai}
\author{Jack M. DiSciacca}
\author{Benjamin L. Lev}
\affiliation{Departments of Applied Physics and Physics and Ginzton Laboratory,  \\ Stanford University, Stanford, CA 94305}

\begin{abstract}
We  demonstrate a novel atom chip trapping system that allows the placement and high-resolution imaging of ultracold atoms within microns from any $\alt$100 $\mu$m-thin,   UHV-compatible material, while also allowing sample exchange with minimal experimental downtime.  The sample  is not connected to the atom chip, allowing rapid exchange  without perturbing the atom chip or laser cooling apparatus.   Exchange of the sample and retrapping of atoms has been performed within a week turnaround, limited only by chamber baking.  Moreover, the decoupling of sample and atom chip provides the ability to independently tune the sample temperature and its position with respect to the trapped ultracold gas, which itself may remain in the focus of a high-resolution imaging system.
As a first demonstration of this new system, we have confined a 700-nK cloud of $8\times10^{4}$ $^{87}$Rb atoms within 100 $\mu$m of a  gold-mirrored 100-$\mu$m-thick silicon substrate.  The substrate was cooled to 35 K without use of a heat shield, while the atom chip, 120~$\mu$m away, remained at room temperature.  Atoms may be imaged and retrapped every 16 s, allowing rapid data collection.
\end{abstract}
\date{\today}
\maketitle

Ultracold gases trapped near cryogenic surfaces using atom chips~\cite{Reichel:2002tr,*Schmiedmayer02,*Zimmermann07} can serve as elements of hybrid quantum systems  for quantum information processing, e.g., by coupling quantum gases to superconducting qubits~\cite{Verdu:2009kt,*Hafezi:2012gz}, or as sensitive, high-resolution, and wide-area probes of electronic current flow~\cite{Schmiedmayer05_Nature,*Schmiedmayer06_APL,*Aigner:2008}, electric ac and patch fields~\cite{Obrecht:2007br,*Bohi:2010ip}, and magnetic domain structure~\cite{Whitlock:2007kj} and dynamics.  Previous experiments have succeeded in trapping and imaging ultracold thermal and quantum gases of alkali atoms around carbon nanotubes~\cite{Gierling:2011cl}, near superconductors~\cite{Nirrengarten:2006eh,*Mukai:2007hh,*Muller:2009gw,*Cano:2011bk} at 4 K,  microns from room-temperature gold wires~\cite{Kruger07}, and within a He dilution refrigerator~\cite{Jessen:2013uf}.

In all of these atom chip experiments, the material to be held in close proximity to the ultracold gas is integrated into the very same substrate forming the atom chip~\cite{Vladan04,Gunther:2005hm,Aigner:2008,Emmert:2009bb,Gierling:2011cl}.  To change samples, one must refabricate the atom chip, before remounting and re-connecting microwires in the UHV chamber.  In cases in which the chip forms a wall of an ultra-compact glass-cell UHV chamber~\cite{Farkas:2010jl,Bohi:2010ip}, rebuilding the entire vacuum apparatus becomes necessary before, finally,  realigning the laser cooling optics.  This poses several challenges to those wishing to trap ultracold atoms near different types of materials in rapid succession, e.g., to test many different hybrid quantum circuit designs or to image the electronic transport in many different cryogenically cooled samples of a strongly correlated or topologically nontrivial material~\cite{Dellabetta:2012fc,*SinucoLeon:2011cq}.  Moreover, attaching the sample directly to the atom chip couples their temperature~\cite{Barclay:2006vq}, preventing one from widely scanning the sample temperature, e.g., to above the superconducting T$_{c}$ for superconducting wires or to below the temperature of water- or LN$_{2}$-cooled normal wires.  Normal microwires rapidly heat\cite{Reichel:2002tr,*Schmiedmayer02,*Zimmermann07,Barclay:2006vq}, modulating the sample temperature during the measurement.

\begin{figure}[b]
\vspace{-6mm}
\begin{center}
\includegraphics[width=0.42\textwidth]{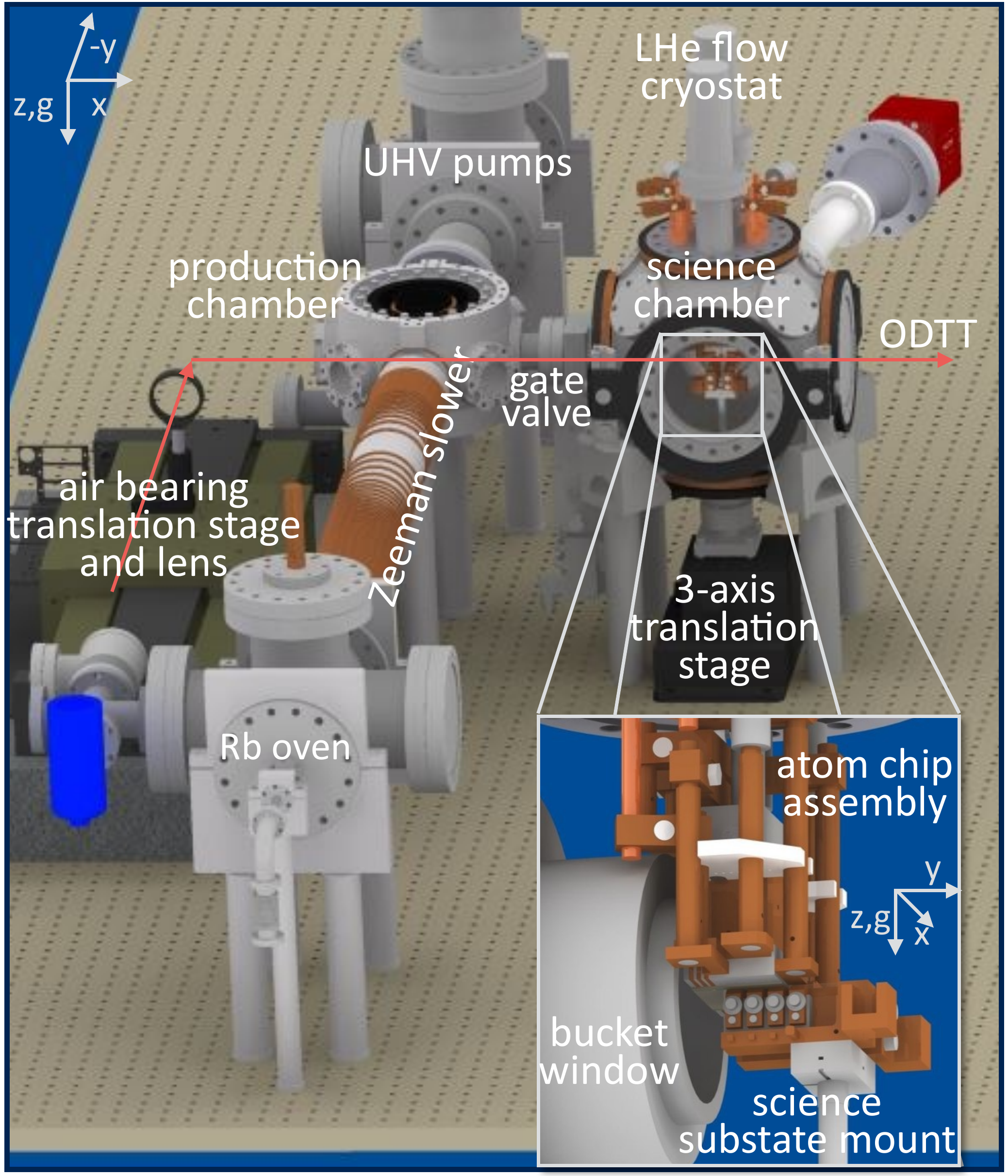}
\caption{\label{Apparatus}Rendering of experimental apparatus for laser cooling and trapping (cooling and imaging optics not shown).  Inset:  Intrachamber atom chip mount and electrical feedthroughs (top) along with the science substrate holder (bottom).  The bucket window (left) allows high-NA lens placement.}
\end{center}
\vspace{-4mm}
\end{figure} 

\begin{figure}[t]
\begin{center}
\includegraphics[width=0.48\textwidth]{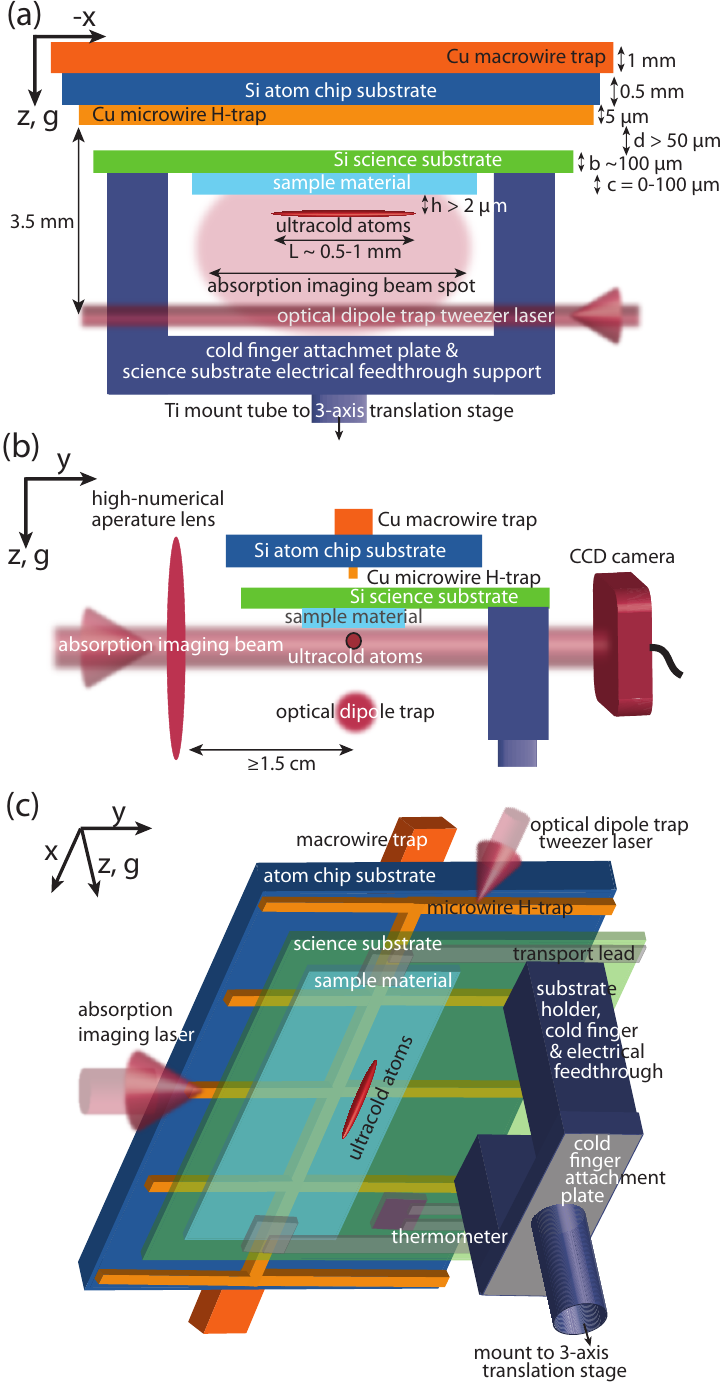}
\caption{\label{ScienceChips}  Schematic of the relative positions of the atoms, sample,  science substrate, holder, and atom chip wires from the viewpoint along the imaging axis (a) and along the optical dipole trap tweezer axis (ODTT) (b). (c) Sketch in perspective of the atom chip with science substrate underneath. }
\end{center}
\vspace{-6mm}
\end{figure} 

\begin{figure}[t]
\begin{center}
\includegraphics[width=0.40\textwidth]{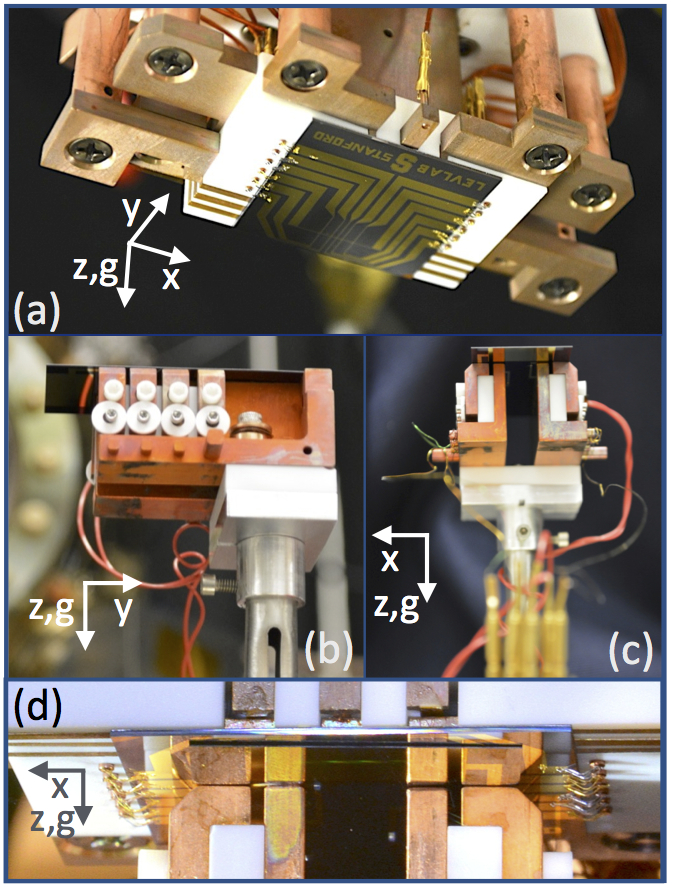}
\caption{\label{ChipPics} (a) Image of atom chip attached to macrowire-macor base chip and chilled-water block.  View of science substrate mount from (b) ODTT viewpoint and (c) imaging beam viewpoint.  (d) Science substrate $d=120$~$\mu$m below atom chip.}
\end{center}
\vspace{-6mm}
\end{figure} 

We report  a new system obviating these problems, allowing us to create, place and image ultracold gases near material surfaces in a manner that allows: 1) rapid exchange of materials without refabrication and reinstallation of a new atom chip and the subsequent realignment of the laser cooling and trapping apparatus;  2) cryogenic cooling and control of material temperature without trapping misalignment due to thermal expansion; 3) ultracold gas production with or without cryogenic surface cooling; 4) flexible placement of contact leads  and thermometers on the material;   5) high-resolution imaging of rapidly created ultracold gases free from trapping-wire-induced ``fragmentation''~\cite{Kruger07}; and 6)  accurate positioning of atoms within microns below any region of a cm$^{2}$  surface.

Specifically, the apparatus shown in Fig.~\ref{Apparatus} can produce ultracold gases of $^{87}$Rb with a $<$10-s repetition rate.  The gas is then transported  33.3~cm  in 4~s with an optical dipole trap tweezer (ODTT)~\cite{Gustavson:2001fw} from the ``production chamber'' through a gate valve to the ``science chamber.''   The decoupling of ultracold gas production from the science chamber, which contains the atom chip and cryogenically cooled sample material and mount, allows us to close the gate valve and break vacuum only in the science chamber when replacing the sample.  The laser cooling optics around the production chamber and its vacuum are not perturbed by the sample exchange procedure or the subsequent science chamber pump-down and high-temperature conditioning.  Only the absorption  imaging optics around the science chamber need be reinstalled.  This simple procedure takes less than week.  We now detail the atom trapping procedure before discussing the design and functionality of the atom chip  and science substrate mount apparatus shown in Figs.~\ref{ScienceChips} and~\ref{ChipPics}.
\begin{figure*}[t]
\begin{center}
 \includegraphics[width=0.95\textwidth]{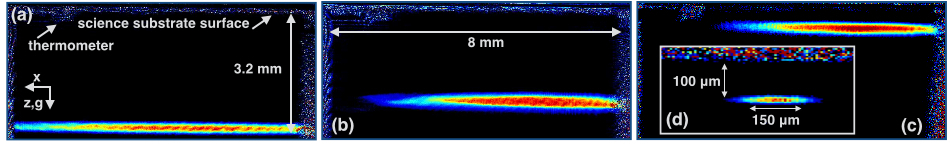}
\caption{\label{AtomPics} Time-of-flight absorption images of: (a) 1.1$\times$$10^{7}$ atoms 2~ms after ODTT release in science chamber; (b) atoms 5 ms after release from macrowire trap; (c) 8$\times$$10^{6}$ atoms at 16 $\mu$K in compressed microwire trap at $h=350$ $\mu$m, 1~ms after release; (d) 8$\times$$10^{4}$ atoms compressed and evaporatively cooled to $[T_{x}, T_{z}] = [470,810]$~nK (mean 700~nK), 1~ms after release.}
\end{center}
\vspace{-4mm}
\end{figure*} 

The design of the production chamber side of the apparatus follows a hybrid magneto-optical evaporation scheme~\cite{Lin:2009it}.  We load a billion $^{87}$Rb atoms into a magneto-optical trap (MOT) from the Zeeman slower within 3--5~s.  A sub-Doppler cooling stage cools the atoms to 21~$\mu$K in 19~ms.  The atoms are then optically pumped for 1.5~ms into the maximally weak-field seeking state $|F,m_{F}\rangle = |2,2\rangle$, where $F$ is the total angular momentum of the $^{87}$Rb atom, before extinguishing all near-resonant lasers and confining the atoms in the magnetic quadrupole trap (MQT) formed by the MOT coils. 
The atoms are compressed by increasing the MQT gradient in 0.8~s to 215~G/cm and subsequent rf-knife evaporation for 3~s cools the gas to 80~$\mu$K.  During rf-evaporation, atoms collect within the trap minimum of a 10-W 1064-nm fiber laser of waist 65~$\mu$m that intersects the MQT 270-$\mu$m below the trap center.  
This ODTT is loaded for 2.1~s before the MQT is reduced to 6 G/cm.  Subsequent evaporation in the ODTT without strong MQT confinement eliminates lifetime-limiting Majorana spin-flip loss. The UHV 2.6$\times10^{-11}$ Torr environment allows us to create a BEC by the rapid reduction of  laser power~\cite{Lin:2009it}.

This laser also  serves as an optical tweezer, enabling the 33.3-cm transfer of ultracold atoms to the atom chip loading zone after MQT turn-off.  We transfer 1.1$\times10^{7}$ atoms at 15 $\mu$K, rather than a BEC, to avoid the 3-body loses found detrimental to ODTT transport with Rb~\cite{Gustavson:2001fw}.  (Atoms may be condensed in the atom chip trap.)  The lens providing the 65-$\mu$m-waist resides on an air bearing translation stage and is smoothly moved to tweezer the ultracold atoms into the science chamber.  The gas is then recaptured from the ODTT by a large volume ``macrowire'' magnetic trap before being compressed and transferred to the atom chip ``microwire'' trap.  Evaporative cooling and trap positioning is then performed.

The macrowire trap is formed by 26~A flowing through a  1~mm$^{2}$ Cu wire combined with a 9.5 G bias field created by an external Helmholtz coils.  This provides 2D confinement, and additional mm-sized wires forming an Ioffe-Pritchard H-trap~\cite{Reichel:2002tr,*Schmiedmayer02,*Zimmermann07} provide confinement along the third dimension $\hat{x}$ in Fig.~\ref{ScienceChips}.  After capturing the atoms at the ODTT position 3.5~mm from the atom chip surface---many beam waists away to minimize heating from scattering light---we change the current and the bias magnetic field within 800~ms  to smoothly pull and compress the atoms toward the sample surface.  Typically 8$\times$$10^{6}$ are transferred from the ODTT  at 14~$\mu$K. The external bias magnetic field is now replaced by the field of 23-A in two parallel 1-mm$^{2}$ wires 2.5-mm on either side of the central  H-trap Cu microwire of the atom chip.  This provides a rapidly quenchable bias field that allows nearly \textit{in-situ}, sub-ms time-of-flight, absorption imaging for atoms in the microwire H-trap.

 We transfer nearly all these atoms within 200~ms from the macrowire trap to the microwire Cu H-trap, whose dimensions are 5-$\mu$m tall and 100~$\mu$m wide (with a 200-nm Au protection layer).  This microwire, and three cross wires for longitudinal confinement, are microfabricated~\cite{Lev2003} on an intrinsic Si substrate.  Finally, the ultracold gas may be rf-evaporatively cooled to BEC~\cite{Reichel:2002tr,*Schmiedmayer02,*Zimmermann07}, or, as presented in Fig.~\ref{AtomPics}, rf-cooled to $<$1~$\mu$K  in a cigar-shaped trap with a population 8$\times10^{4}$.  The atoms are positioned 100(5)~$\mu$m from the substrate with $\hat{x},\hat{z}$ trapping frequencies  $2\pi\times[20, 200]$~Hz using  2.5 A in the central wire and 0.5 A in side wires.   The atoms may then be maneuvered with trapping fields to $h\agt2$~$\mu$m below the cryogenically cooled sample material of thickness $c$, typically 0--100 $\mu$m.  The sample substrate of thickness $b=100$ $\mu$m is held $d = 120$~$\mu$m from the atom chip microwires in Fig.~\ref{AtomPics}d ($d\agt 50$ $\mu$m possible), with the atoms held $ a= d + b +h + c = 320$ $\mu$m from the surface of the  Cu microwires.  This eliminates condensate fragmentation from the disordered trapping wire itself, which has been shown to be detrimental $<$100~$\mu$m from the wire~\cite{Kruger07}, while allowing a small $h$, limited only by Casimir-Polder potentials~\cite{Vladan04}.  At the minimum achievable $a \approx 200$~$\mu$m, a transverse trap frequency of $2\pi \times3.8$~kHz could be obtained with a 3.5-A central microwire current.

Moreover, small $h$'s may be achieved without moving the ultracold atomic gas with respect to the UHV chamber, allowing the placement of the gas within the small depth of field of a rigid high-numerical aperture (NA = 0.5) lens system.   The inset in Fig.~\ref{Apparatus} shows the atom chip assembly with  nearby bucket window for outside-vacuum high-NA lens placement---within 1.5~cm of the ultracold atoms---which could provide 1-$\mu$m imaging resolution~\cite{Hung:2011hy}.

\begin{figure}[t]
\begin{center}
\includegraphics[width=0.47\textwidth]{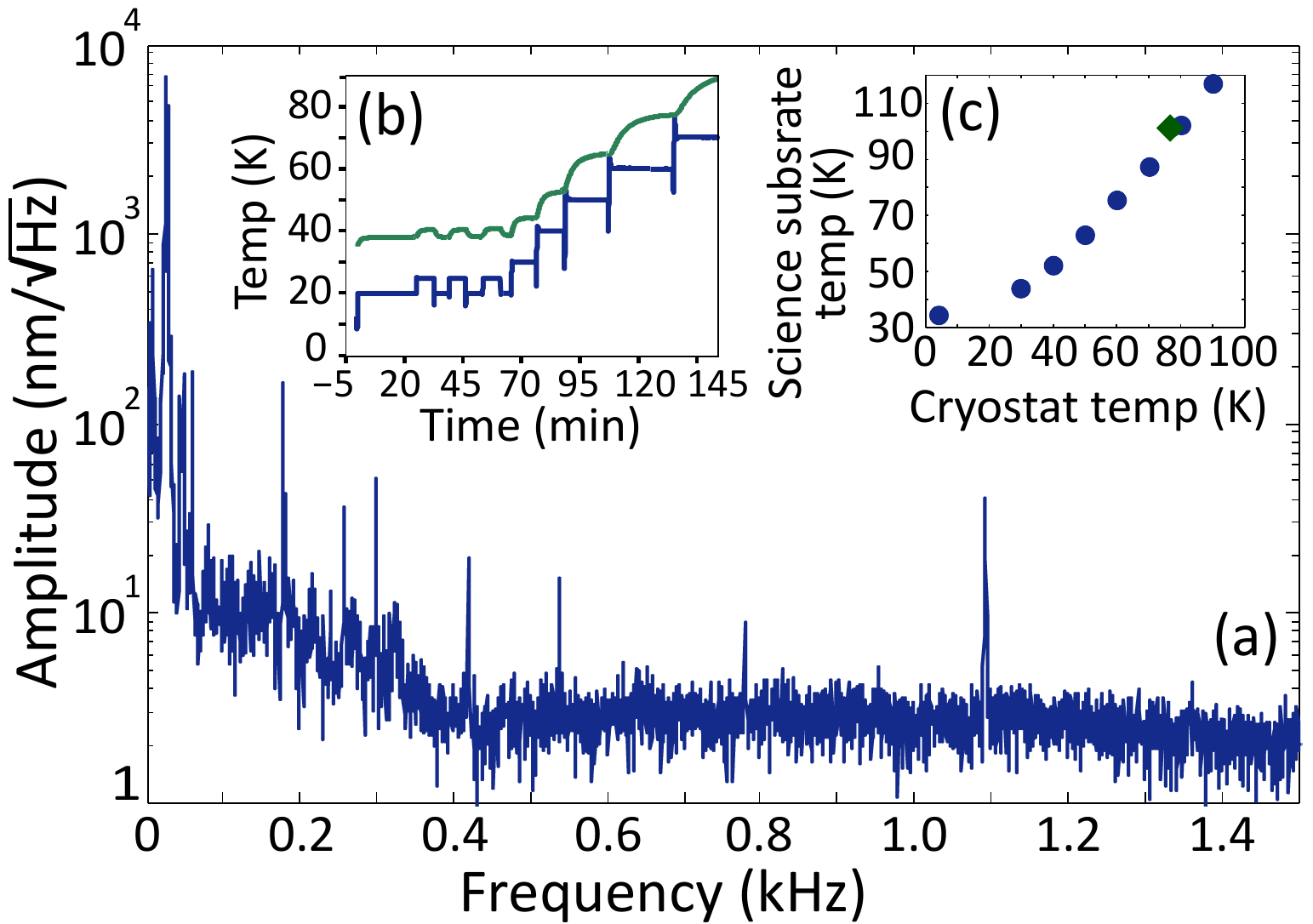}
\caption{\label{TempVib}(a) Vibration amplitude of cantilevered science  substrate measured with Mach-Zehnder interferometer. (b) Temperature step-response of  science  substrate (top, green) vs cold finger (bottom, blue). (c) Steady-state  science  substrate  temperature using LHe (dots) and LN$_{2}$ (diamond).}
\end{center}
\vspace{-4mm}
\end{figure}

Rather than scanning the trapped gas  over wide areas above a sample using magnetic fields, as done in Refs. 3 and which would require the coordinated  movement of the high-NA lens system, our system is able to move the sample material itself:  As shown in Figs.~1--3, the science material sample may be glued or fabricated onto the substrate glued to the Cu mount held in place by a thin Ti tube---for low thermal conductivity---and attached to a room-temperature bellows whose position is controlled by a 3-axis translation stage external to the vacuum chamber.  The cantilevered science sample substrate vibrates no more than 150~nm RMS, well below our imaging resolution, as measured by an \textit{in-situ} Mach-Zehnder interferometer retroreflecting a laser off the Au-mirrored face of the substrate.  Figure~\ref{TempVib} shows the vibration spectral density, dominated by a mechanical resonance at 26~Hz.

The science sample mount is heat-sunk via flexible Cu braids to a  liquid-He flow cryostat cold finger.  Upon cool-down, the atom chip position---and the trapped ultracold gas---remains fixed and in the focus of the lens system, while the $\sim$300 $\mu$m thermal contraction of the sample mount is easily compensated by the 3-axis translation stage.  The current experiment has a science substrate with a simple gold mirror as a sample material, and the gold mirror is lithographically patterned to provide electrical contacts to two silicon diode thermometers, each glued onto the substrate in a manner unobtrusive to the  ODTT or imaging beam optical access; one is near the tip of the cantilevered sample, the other below the Cu substrate mount.  We measure no temperature gradient between the substrate thermometers, within their intrinsic error.  Moreover, Fig.~\ref{TempVib}  shows only a few-min equilibration time delay between the cold finger and substrate during cooling.

Without a heat shield, our sample tip base temperature is 35 K when the cold finger is 4.3~K or  100 K when cooling with 77 K LN$_{2}$. With a radiation heat shield---fabricated though not yet installed---we predict that sub-10 K temperatures are achievable with the current cryostat's cooling power; $\sim$4 K may be achievable  by cooling the atom chip with LN$_{2}$ rather than chilled water.  Despite these thermometers---and their electrical contacts---we achieve a room-temperature pressure of $2\times10^{-10}$ Torr (trap lifetime $\sim$10 s, which will soon be improved using larger ion pumps)  and  $5\times10^{-11}$ (trap lifetime $\sim$20 s) under LHe cooling, sufficient for BEC production.   A cold-finger thermometer and heater provide temperature control to  $<$0.1~K at the substrate tip.

The demonstrated instrument serves as an alternative to single-use multilayer chips\cite{Bohi:2009di,*Trinker:2008iy,*Chuang:2011kk} and paves the way for future microscopy experiments using, e.g., ultracold Rydberg atoms to probe surface patch potentials\cite{Hattermann:2012ud,*Carter:2012jb} or to couple to hybrid quantum circuits\cite{Verdu:2009kt,*Hafezi:2012gz}, or to use ultracold atoms and BECs to rapidly image the transport in samples of strongly correlated or topologically protected materials\cite{Dellabetta:2012fc,*SinucoLeon:2011cq}.

We thank  M.~Tekant, K.~Baumann, A.~Koll{\'{a}}r, M. Armen, S.-H.~Youn, V.~Redmon, B.~Kasch, and U.~Ray for experimental assistance and P.~Abbamonte and L.~Cooper for  cryogenics advice.  We acknowledge generous funding from the U.S. Department of Energy, Office of Basic Energy Sciences, Division of Materials Sciences and Engineering under award \#DE-SC0001823 for construction of the instrument and support of B.L.L.  We also thank  the Gordon and Betty Moore Foundation through grant GBMF3502 for support of R.W.T.~and J.M.D.  R.A.L. acknowledges support from the Hertz Foundation.
%


%

\end{document}